\documentclass[lettersize,journal]{IEEEtran}
\usepackage{amsmath,amsfonts}
\usepackage{algorithmic}
\usepackage{algorithm}
\usepackage{array}
\usepackage[caption=false,font=normalsize,labelfont=sf,textfont=sf]{subfig}
\usepackage{textcomp}
\usepackage{stfloats}
\usepackage{url}
\usepackage{verbatim}
\usepackage{graphicx}
\usepackage{color}
\usepackage{cite}
\usepackage{multirow} 
\usepackage{makecell}

\begin{document}

\title{Shared Sky, Shared Spectrum: Coordinated Satellite-5G Networks for Low-Altitude Economy}

\author{\IEEEauthorblockN{Yanmin Wang, Wei Feng, \emph{Senior Member, IEEE}, Yunfei Chen, \emph{Fellow, IEEE}, Baoquan Ren, Qingqing Wu, \emph{Senior Member, IEEE}, and Cheng-Xiang Wang, \emph{Fellow, IEEE}}	
	\thanks{This work was supported in part by the National Natural Science Foundation of China under Grant 62541104, Grant 62425110, and Grant U22A2002, in part by the Major Science and Technology Project of Jiangsu Province under Grant BG2025039, in part by the Research Fund of National Mobile Communications Research Laboratory, Southeast University, under Grant 2026A05, and in part by the FAW Jiefang Automotive Co., Ltd.}
	\thanks{
		Y. Wang is with the School of Information Engineering, Minzu University of China, Beijing 100081, China~(email: wangyanmin@muc.edu.cn).
		
		W. Feng (corresponding author) is with the Department of Electronic Engineering, State Key Laboratory of Space Network and Communications, Tsinghua University, Beijing 100084,
		China~(email: fengwei@tsinghua.edu.cn). 
		
		Y. Chen is with the Department of Engineering, University of Durham, Durham DH1 3LE, U.K. (email: yunfei.chen@durham.ac.uk).
		
		B. Ren is with the China Electronic System Engineering Company, Beijing, China~(email: renbq88@126.com).
		
		Q. Wu is with the Department of Electronic Engineering, Shanghai Jiao Tong University, Shanghai 200240, China (email: qingqingwu@sjtu.edu.cn).
		
		C.-X. Wang is with the National Mobile Communications Research Laboratory, School of Information Science and Engineering, Southeast University, Nanjing 210096, China, and also with the Purple Mountain Laboratories, Nanjing 211111, China (email: chxwang@seu.edu.cn).
	}
}



\maketitle

\begin{abstract}
Driven by both technological development and practical demands, the low-altitude economy relying on low-altitude aircrafts (LAAs) is booming. However, neither satellites nor terrestrial fifth-generation (5G) networks alone can effectively satisfy the communication requirements for ubiquitous low-altitude coverage. 
While full integration of satellites and 5G networks offers theoretical benefits, 
the divergent legacies and disparate propagation characteristics pose significant challenges for its rapid deployment. 
As a more economical and immediately-viable alternative, this paper investigates partially-integrated networks where satellites and 5G systems operate with coarse synchronization yet achieve coordinated spectrum sharing, pooling their capabilities to jointly serve LAAs. Leveraging the inherent position-awareness of LAAs, we propose a framework for joint time-frequency spectrum sharing with an adaptive synchronization time scale, where only large-scale channel state information (CSI) is required. To avoid solving the NP-hard optimization problem directly, link-feature-aided clustering is employed following a divide-and-conquer strategy. The proposed framework achieves substantial performance gains with low overhead and complexity, enabling swift advancement of low-altitude applications while paving the way for future integrated satellite-terrestrial network evolution.
\end{abstract}


\section{Introduction}
Low-altitude economy is rising rapidly worldwide~\cite{LAE_twc_2025, LAE_tcom_2025}, enabling production, delivery, emergency rescue, tourism, and exploration through the deployment of low-altitude aircrafts (LAAs) across urban, rural, mountainous, inshore, offshore, and other remote areas~\cite{LAE_twc_2025, fw_jsac_2025}. This burgeoning sector demands ubiquitous and reliable communication coverage to ensure both flight safety and mission success. However, existing terrestrial fifth-generation (5G) networks were originally designed for ground users, and extending coverage to low altitudes faces significant challenges from complex interference and blind spots~\cite{5G_book}. Conversely, 
while satellite systems can offer wide-area coverage, they
suffer from inherent limitations in serving low-altitude operations, including  
relatively limited data rates and large path loss and latency. 
Consequently, neither satellites nor terrestrial 5G networks alone can effectively satisfy the stringent communication requirements for ubiquitous low-altitude coverage.

Hybrid satellite-terrestrial networks present a promising solution by leveraging the complementary capabilities of both
the terrestrial 5G and satellite systems~\cite{xx2}. 
To fully exploit the advantages of hybrid networks,
efficient spectrum utilization should be pursued to tackle with the persistent challenge of spectrum scarcity
by adopting various state-of-the-art techniques~\cite{revision_add, Chinacom_2025,r_twc_2022,10678835}.
In particular, spectrum sharing between satellite and terrestrial links provides an effective option 
due to the rich opportunities brought by the spatially-scattered distribution of users including LAAs 
in a hybrid network~\cite{Chinacom_2025,r_twc_2022,10678835}.
In both satellite and 5G networks,  
the spectrum is usually utilized in terms of time-slotted carriers based on time division multiple access (TDMA)
for efficiency and flexibility~\cite{5G_book,r_JSAC_2021,ref_DVB_2024, WCL_2025}.
With potential for higher performance gain, dynamic spectrum sharing in both the frequency and time domains 
is much preferred~\cite{xx2, 10678835, r_twc_2022, r_JSAC_2021, WCL_2025}.
Unfortunately, along the way of regular spectrum sharing in the time domain,
time-slot-level synchronization is taken as a fundamental prerequisite~\cite{xx2,10678835}. 
It is too stringent to be achieved in a hybrid network anticipated to 
align with the urgent demands of emerging low-altitude applications in the short run~\cite{10678835, r_JSAC_2021,5G_book,ref_DVB_2024}.

Although the satellite-terrestrial network has been extensively addressed,
achieving full integration between satellites and terrestrial 5G networks encounters substantial practical obstacles.
In a fully-integrated network,
air interfaces with unified design, e.g., consistent waveforms, frame structures, and interaction protocols,
are expected to be deployed in the satellite and terrestrial components~\cite{xx1}.
With both divergent legacies and disparate propagation characteristics in the satellite and terrestrial 5G networks,
the unification of air interfaces is far from either a short-term or an easy-to-accomplish task~\cite{5G_book,ref_DVB_2024}.
For fast-emerging low-altitude applications,
the hybrid network with partially-integrated satellites and terrestrial 5G systems
is a more pragmatic and immediately-deployable alternative.
Instead of relying on unified air interfaces,
a partially-integrated network focuses on aggregating serving capabilities 
of the satellite and terrestrial 5G components, e.g., via coordinated link scheduling and resource allocation.
Under partial integration,
time-slot-level satellite-terrestrial synchronization is rather challenging for the hybrid network~\cite{10678835, r_JSAC_2021,5G_book,ref_DVB_2024}.
With unavoidable frequent negotiating interactions and excessive system overhead,
it will be prohibitively-complex and resource-intensive~\cite{xx2}.

To support the urgent demands of low-altitude applications,
this paper investigates partially-integrated networks where satellites and 5G systems cooperate with coarse synchronization.
Joint time-frequency satellite-terrestrial spectrum sharing is addressed
to fill the gap in previous research.
Under coarse synchronization conditions, existing studies have predominantly concentrated on frequency-domain spectrum sharing alone~\cite{r_twc_2022,10678835}.
Leveraging the inherent position-awareness of LAAs, we propose a novel framework that enables efficient spectrum coordination without requiring stringent synchronization.
By offering a balanced trade-off between performance enhancement and implementation feasibility,
it facilitates rapid advancement of low-altitude applications while establishing foundations for future evolution
to fully-integrated satellite-terrestrial networks.

\section{Opportunities and Challenges for Spectrum Sharing in Coordinated Networks Serving LAAs}
\subsection{Opportunities}
In hybrid satellite-5G networks, both satellites and terrestrial base stations (TBSs) can be utilized to provide communication access for LAAs. Satellite and 5G links coexist to serve not only LAAs but also large numbers of terrestrial users (TUs), including both 5G and satellite terminals. Due to the high altitude of satellites, the coverage area of a satellite beam is usually much larger than that of a TBS~\cite{r_JSAC_2021}. Furthermore, partial or full frequency reuse is typically adopted among different TBSs in terrestrial networks~\cite{5G_book}. Consequently, if satellite-terrestrial spectrum sharing is implemented, one satellite link tends to interfere with a large number of terrestrial links. Nevertheless, owing to the spatially-scattered distribution of LAAs and existing satellite/5G TUs, substantial opportunities still exist for satellite-terrestrial spectrum sharing.

\begin{figure*} [t]
	\centering
	\includegraphics[width=16.5cm]{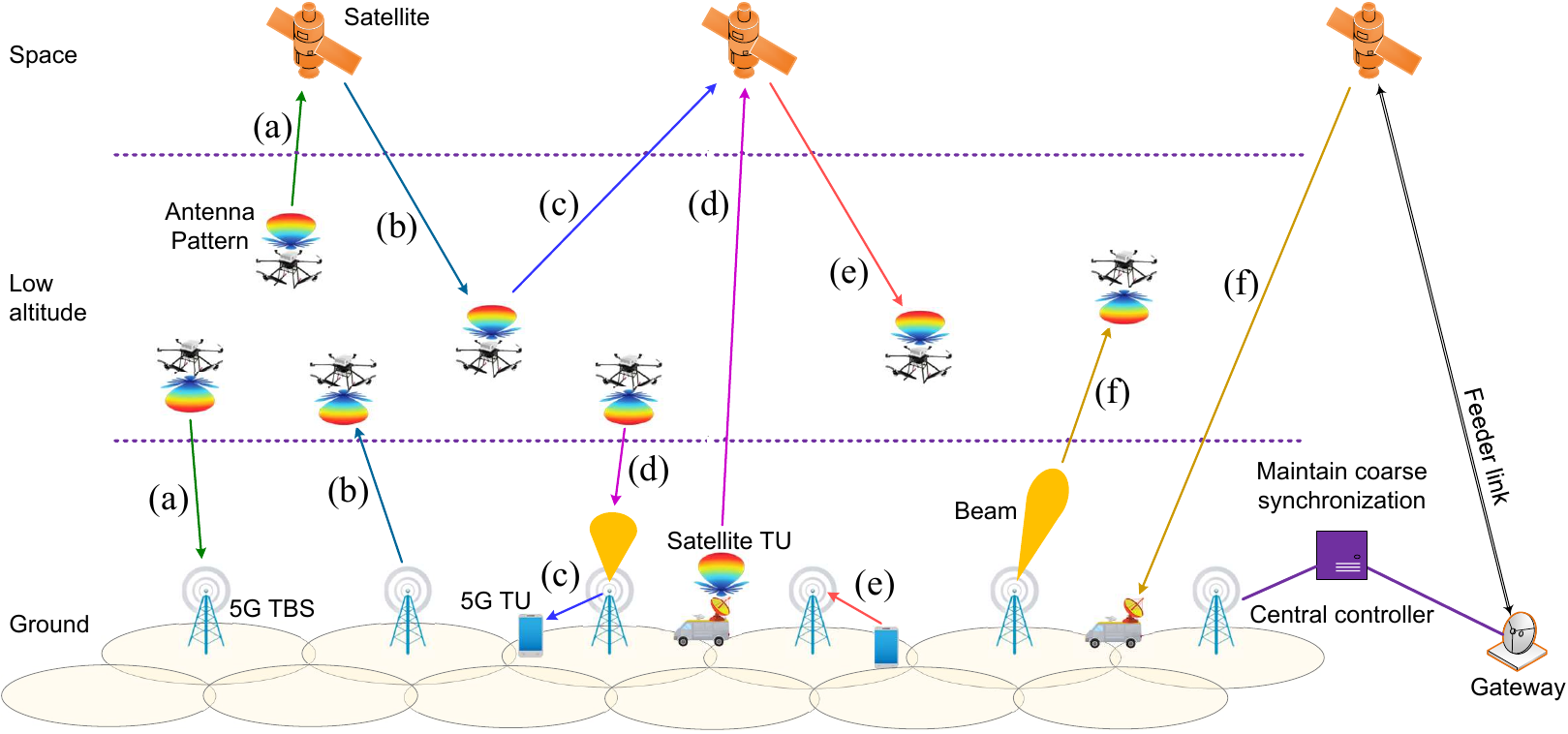}
	\caption{Opportunistic satellite-terrestrial spectrum sharing for  
			(a) LAA-satellite uplink and LAA-TBS uplink,
			(b) satellite-LAA downlink and TBS-LAA downlink, 
			(c) LAA-satellite uplink and TBS-TU downlink,
			(d) TU-satellite uplink and LAA-TBS uplink,
			(e) satellite-LAA downlink and TU-TBS uplink,
			and (f) satellite-TU downlink and TBS-LAA downlink.}
	\label{fig_sharing_opportunity}
\end{figure*}

Fig.~\ref{fig_sharing_opportunity} illustrates several typical hybrid networking scenarios where spectrum sharing can be opportunistically implemented between satellite and 5G links. Specifically, (a) and (b) depict spectrum sharing opportunities between satellite and 5G links both serving LAAs, while (c)--(f) illustrate those between links serving LAAs and satellite/5G TUs, respectively. Links of the same color in Fig.~\ref{fig_sharing_opportunity} indicate that satellite and 5G links can utilize the same carriers simultaneously.

The spatially-scattered distribution of LAAs and satellite/5G TUs, combined with the directional transmission and reception capabilities of TBSs and LAAs, create opportunistic spatial separations, i.e., low-interference opportunities, between satellite and 5G links. Whenever such spatial isolation exists, spectrum sharing between satellite and 5G links becomes viable. These isolation conditions can be exploited for inter-link interference mitigation through cooperative link scheduling and transmission optimization techniques such as power control~\cite{Chinacom_2025,r_twc_2022,10678835}, thereby achieving more efficient spectrum utilization and 
capability aggregation between the satellite and 5G systems.

\subsection{Challenges}
For joint time-frequency spectrum sharing,
time synchronization among links is a fundamental prerequisite~\cite{r_JSAC_2021, ref_DVB_2024}.
With time-slot-level fine synchronization,
inter-link interference caused by spectrum sharing can be dynamically evaluated and mitigated. 
However, achieving millisecond-level synchronization between satellite and 5G links is quite challenging~\cite{5G_book,10678835, ref_DVB_2024}, mainly due to two factors.

\begin{figure*} [t]
	\centering
	\includegraphics[width=17cm]{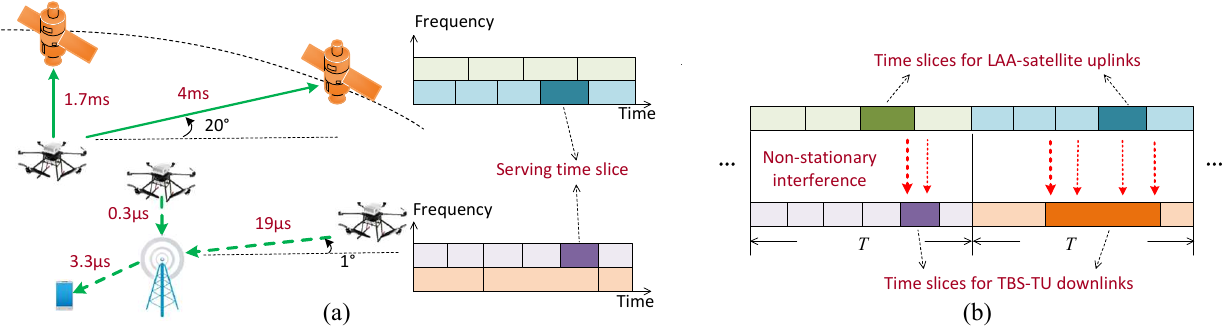}
	\caption{(a) Propagation delays and serving time slices for satellite and terrestrial links, and (b) coarse satellite-terrestrial synchronization, in coordinated satellite-5G networks.}
	\label{fig_sync_challenge}
\end{figure*}

\textbf{First}, satellite and terrestrial links experience vastly different propagation delays owing to their distinct transmission distances. As illustrated in the left part of Fig.~\ref{fig_sync_challenge}(a), the propagation delay from an LAA at $100$m to a satellite orbiting at $500$km altitude varies rapidly between $1.7$ms and $4$ms. In contrast, the delay from a 5G TBS to the LAA ranges from approximately $0.3$ microseconds ($\mu$s) to $19$$\mu$s, while the delay from a TBS to a TU is at most about $3.3$$\mu$s when the coverage radius is $1$km.

\textbf{Second}, 
the durations of serving time slice allocated to satellite and terrestrial links can be significantly divergent
due to distinctive time slot division and allocation strategies. 
For channel-adapted transmission and quality of service (QoS) guarantee, different time division schemes are usually adopted for satellite and terrestrial links, where multiple basic time slots may be aggregated to form a serving time slice~\cite{5G_book,ref_DVB_2024}. 
As shown in the right part of Fig.~\ref{fig_sync_challenge}(a), 
this leads to divergent serving time slice durations across the links. 
Although it is based on current standards for terrestrial and satellite networks, the figure is expected to reflect the technical tendency in future networks as well.

Given these divergent propagation delays and serving time slice durations, 
time-slot-level fine time synchronization between satellite and terrestrial links becomes difficult to achieve. 
Although timing advance can compensate for delay differences, 
the rapid variation of satellite link delay makes it challenging to implement 
for fine satellite-terrestrial synchronization in practical coordinated networks. 
It can introduce heavy signaling overhead, high implementation complexity, and non-negligible spectrum waste~\cite{xx2}.
Particularly, fine satellite-terrestrial synchronization
may remain as a complex and resource-intensive task
even in a fully-integrated network because of the largely-different propagation delays of satellite and terrestrial links~\cite{5G_book, 10678835, xx1, xx2}.
As an alternative, coarse satellite-terrestrial synchronization over a larger time scale $T$ shown in Fig.~\ref{fig_sync_challenge}(b)
can be considered~\cite{10678835, ref_TWC_2025_submitted}.

\section{Cooperative Spectrum Sharing under Coarse Satellite-Terrestrial Synchronization}
\subsection{Fundamental Idea of Coarse Satellite-Terrestrial Synchronization}
To ensure the viability of joint time-frequency satellite-terrestrial spectrum sharing in practical implementations, it is crucial to maintain overhead and complexity at appropriate levels~\cite{Chinacom_2025,r_twc_2022,10678835}. To circumvent the challenges of fine time synchronization discussed above, we can rely on coarse satellite-terrestrial synchronization for spectrum sharing~\cite{10678835, ref_TWC_2025_submitted}. Fig.~\ref{fig_sync_challenge}(b) illustrates the fundamental idea using a time-slotted carrier shared by LAA-satellite uplinks and TBS-TU downlinks as an example, like the case (c) in Fig.~\ref{fig_sharing_opportunity}.

For the implementation of TDMA in the carrier, fine synchronization over time slots is required among satellite links and among terrestrial links, respectively. In contrast, synchronization between satellite and terrestrial links is achieved only at a coarse time scale $T$ for spectrum sharing~\cite{10678835}. The durations of serving time slices for satellite and terrestrial links are determined independently, with their boundaries aligned at the coarse time scale $T$, just as shown in Fig.~\ref{fig_sync_challenge}(b).

Consistent with coarse satellite-terrestrial synchronization, satellite and terrestrial links can be cooperatively scheduled and optimized at the same time scale $T$ for spectrum sharing~\cite{10678835}. Inter-link interference estimation and mitigation also adopt $T$ as the basic reference time scale. Accordingly, the serving time of the hybrid network can be divided into periodic intervals with duration $T$ across all carriers. As shown in Fig.~\ref{fig_sync_challenge}(b), in each time interval $T$, one satellite link may interfere with or be interfered by different 5G links scheduled in adjacent time slices of the same carrier, and vice versa. This leads to non-stationary interference for any given satellite or 5G link, which requires special attention to satisfy the QoS requirements of LAAs and TUs.

For satellite-terrestrial cooperation at the coarse time scale $T$, it is more viable to utilize large-scale rather than instantaneous channel state information (CSI)~\cite{Chinacom_2025}. 
It is straightforward to note that
the performance gain achieved by spectrum sharing tends to decrease as $T$ increases. Nevertheless, larger time scales imply lower overhead and complexity. To balance performance requirements and practical constraints, the value of $T$ should be selected according to 
characteristics of the specific hybrid network scenario.

\subsection{Complicated Inter-Link Coupling and A Possible Way Out}
Satellite-terrestrial spectrum sharing can be realized through cooperative link scheduling and transmission optimization, such as beamforming and power control~\cite{xx2, Chinacom_2025,r_twc_2022,10678835}. When multiple satellite beams or TBSs are available for an LAA or a TU simultaneously, satellite, beam, or TBS selection can also be utilized. Correspondingly, the spectrum sharing optimization can be formulated as a mixed integer programming (MIP) problem. Owing to the large coverage area of a satellite beam and frequency reuse among TBSs, one satellite link could be coupled with multiple terrestrial links by interference. As indicated by Fig.~\ref{fig_sync_challenge}(b), coarse satellite-terrestrial time synchronization further complicates this coupling. 
Satellite and terrestrial links sharing the same carrier in a specific time interval $T$ could be all mutually coupled via inter-link interference under coarse synchronization. 
When embedded in the objective functions and/or constraints indicating the network performance or QoS for LAAs and TUs, 
as is the usual case,
such complicated inter-link coupling renders the spectrum sharing problem NP-hard~\cite{10678835}.

In satellite-terrestrial spectrum sharing, cooperative link scheduling is a main contributor to the high complexity of the formulated problem~\cite{10678835}. Specifically, with coarse satellite-terrestrial synchronization, scheduling orders of satellite and terrestrial links sharing the same carrier in a time interval $T$ can be arranged independently. It promises greater flexibility in providing QoS-guaranteed services for LAAs and TUs, such as low-latency data transmission. In this case, interference may occur between any pair of satellite and terrestrial links sharing the same carrier in a time interval $T$. Accordingly, links sharing the same carrier in a time interval $T$ actually form two clusters: a satellite link cluster and a terrestrial one. For spectrum sharing, inter-link interference needs to be estimated and mitigated based on link clusters rather than individual links~\cite{10678835}. Following this observation, cooperative scheduling of satellite and terrestrial links can be implemented efficiently via link clustering.

\section{Link-Feature-Aided Satellite-Terrestrial Spectrum Sharing with Adaptive Cooperation Time Scale}

\subsection{Link Clustering based on Interference Similarity}
\begin{figure*} [t]
	\centering
	\includegraphics[width=16cm]{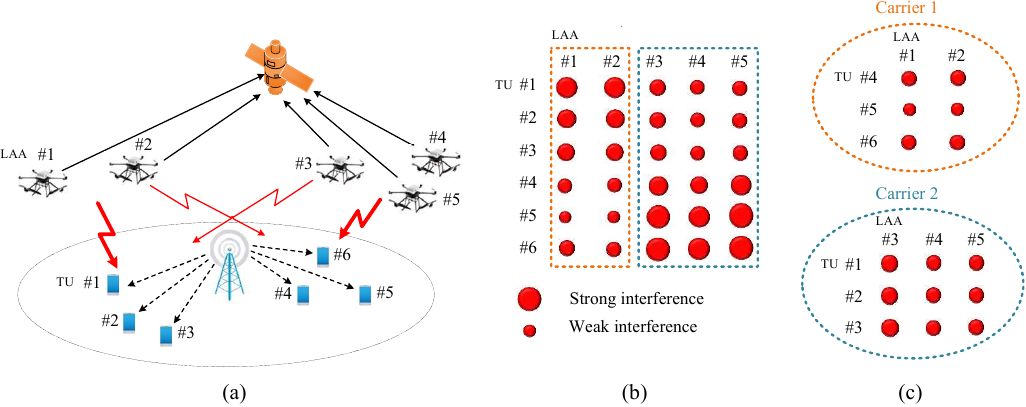}
	\caption{Illustration of interference-similarity-based link clustering for cooperative link scheduling.}
	\label{fig_intf_profile}
\end{figure*}

As pointed out in Section III, with coarse satellite-terrestrial synchronization, cooperative link scheduling for joint time-frequency spectrum sharing can be accomplished via link clustering. This requires that interference between any pair of satellite and terrestrial links in two link clusters sharing the same carrier in each time interval $T$ remains low. In particular, to fully exploit opportunistic separations between satellite and terrestrial links, link clustering can be performed efficiently based on inter-link interference similarity~\cite{10678835, ref_TWC_2025_submitted}. That is satellite or terrestrial links that cause or suffer similar interference to or from terrestrial or satellite links are clustered together. The fundamental idea of interference-similarity-based link clustering for cooperative link scheduling is illustrated in Fig.~\ref{fig_intf_profile}.

As shown in Fig.~\ref{fig_intf_profile}(a), it is assumed that LAA-satellite uplinks and TBS-TU downlinks share the same group of carriers. Due to the downtilt of antenna arrays at TBSs toward TUs and the much larger distances from TBSs to satellites, interference from TBS-TU links to LAA-satellite links can be ignored. Thus, only interference leaked from the back lobes of antenna arrays at LAAs to TBS-TU links is considered. The strength of interference from LAAs to TUs is presented in Fig.~\ref{fig_intf_profile}(b). It can be observed that based on interference similarity, the LAA-satellite links for LAAs $1$ and $2$ can be clustered together, as can those for LAAs $3$, $4$, and $5$. As shown in Fig.~\ref{fig_intf_profile}(c), by fitting the two LAA-satellite link clusters into two carriers, the TBS-TU links can be opportunistically scheduled.

On one hand, by clustering satellite and terrestrial links with similar inter-link interference together, opportunistic separations between individual links are preserved in link clusters. On the other hand, it keeps interference suffered by each link as stationary as possible at a low level in each time interval $T$ after mapping the link clusters to time-slotted carriers. As a result, interference between satellite and terrestrial links is effectively mitigated through clustering-based link scheduling. By virtue of these opportunistic inter-link separations, the obtained link clusters can efficiently fit available time-slotted carriers in each time interval $T$. Furthermore, by applying cooperative transmission optimization to the links in each carrier, joint time-frequency spectrum sharing with coarse satellite-terrestrial synchronization can be realized.

Note that for QoS guarantee, the serving time required by different LAAs or TUs in a time interval $T$ can be quite distinct. Thus, the duration of the serving time slice allocated to each satellite or terrestrial link should be tailored according to QoS requirements. Correspondingly, the number of satellite or terrestrial links in each link cluster needs to be adapted to QoS and may vary across clusters. As shown in Fig.~\ref{fig_intf_profile}(b) and Fig.~\ref{fig_intf_profile}(c), there are $2$ and $3$ LAA-satellite links in the two link clusters, respectively. This is an important consideration for link clustering in addition to interference mitigation.

\subsection{Link Feature Sketching for Interference-Similarity-based Link Clustering}
For efficient implementation of link clustering, a promising approach is to use classic clustering algorithms, such as K-means~\cite{ref_TWC_2025_submitted}. Specifically, feature vectors can be designed for satellite and terrestrial links to evaluate the similarity of interference suffered or caused by them. The smaller the distance between feature vectors for two satellite or terrestrial links, the more similar the links are in terms of inter-link interference. Correspondingly, to accomplish link clustering, the key lies in how to evaluate the similarity of interference between satellite and terrestrial links in spectrum sharing. 
Targeting at optimizing the network performance while guaranteeing QoS for LAAs and TUs, the similarity should not be evaluated merely based on interference strength. Instead, it is more appropriate to evaluate it based on the impact of interference on network performance and QoS~\cite{10678835, ref_TWC_2025_submitted}. Further, the impact of interference can be measured by the resulting link quality degradation, and feature vectors for the links can be designed accordingly.

\begin{figure} [t]
	\centering
	\includegraphics[width=8.5cm]{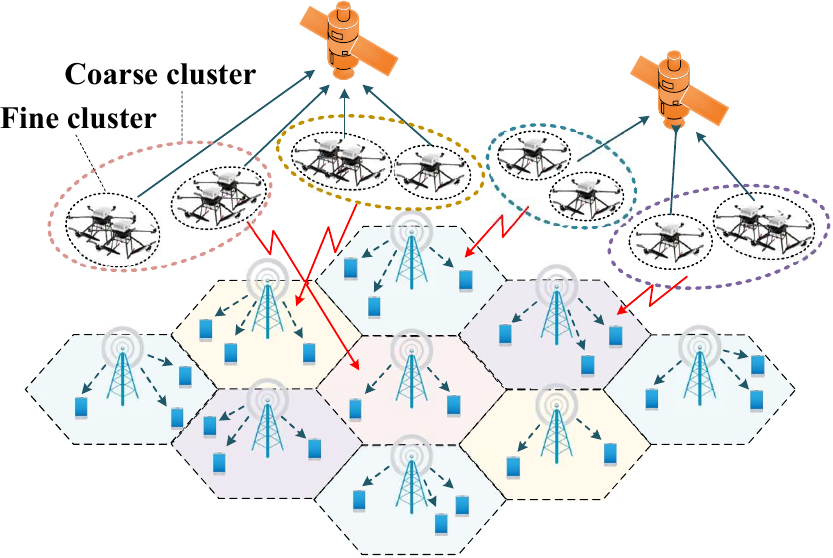}
	\caption{A hybrid network with satellite-terrestrial spectrum sharing between LAA-satellite and TBS-TU links.}
	\label{fig_network_example}
\end{figure}

Consider the hybrid network shown in Fig.~\ref{fig_network_example}.
Suppose $U$ LAA-satellite links and $N$ TBS-TU links share $K$ time-slotted carriers opportunistically in a time interval $T$.
As explained above, 
only interference leaked from back lobes of antenna arrays at LAAs to TBS-TU links needs to be considered.
For the clustering of LAA-satellite links, feature vectors can be designed for them as 
\begin{equation}\label{eq1}
	\begin{split}
		\mathcal{F}_u = & [\, \mathcal{R}_1 ( \mathbb{T}_1, \mathbb{T}'_u \,|\, \mathbb{H} ), \mathcal{R}_2 ( \mathbb{T}_2, \mathbb{T}'_u \,|\, \mathbb{H} ), \,\,\, ..., \\
		& \,\,\,\,\,\,\,\,\,\,\,\,\,\,\,\,\,\,\,\,\,\,\,\,\,\,\,\,\,\,\,\,\,\,\,\,\,\,\,\,\,\,\,\,\,\,\,\,\,\,\,\, \mathcal{R}_N ( \mathbb{T}_N, \mathbb{T}'_u \,|\, \mathbb{H} ) \, ],
	\end{split}
\end{equation}
where $\mathcal{F}_u$ is the feature vector for the LAA-satellite link serving LAA $u$, $u=1,...,U$.
$\mathbb{T}_n$, $n=1,...,N$, denote the transmission parameters of TBSs for TUs, such as transmit power and beamforming vectors,
$\mathbb{T}'_u$, $u=1,...,U$, are those of the LAAs,
$\mathbb{H}$ represents large-scale CSI of all links in the time interval, including that of the interference links from LAAs to TUs,
and $\mathcal{R}_n ( \mathbb{T}_n, \mathbb{T}'_u \,|\, \mathbb{H} )$, $n=1,...,N$, denote the degraded quality of 
all TBS-TU links, e.g., average transmission rate,
with the interference caused by $\mathbb{T}'_u$ under $\mathbb{H}$.
Further, the distance between any two feature vectors $\mathcal{F}_{u_1}$ and $\mathcal{F}_{u_2}$
can be defined as
\begin{equation}\label{eq2}
	\mathcal{D}_{u_1,u_2} = || \mathcal{F}_{u_1} - \mathcal{F}_{u_2} ||_1,
\end{equation}
where $||\cdot||_1$ denotes the $L_1$ norm~\cite{10678835,ref_TWC_2025_submitted}.

The feature vectors given by~(\ref{eq1}) can actually be seen as
a kind of distributed representation of the impact of inter-link interference.
Matching with the optimization target, such as sum rate maximization and energy consumption minimization,
$\mathcal{R}_n ( \mathbb{T}_n, \mathbb{T}'_u \,|\, \mathbb{H} )$ in (\ref{eq1}) 
can be expressed in different forms~\cite{10678835,ref_TWC_2025_submitted}.
Under large-scale CSI represented by $\mathbb{H}$,
$\mathcal{R}_n ( \mathbb{T}_n, \mathbb{T}'_u \,|\, \mathbb{H} )$ has to be calculated 
via expectation with respect to the random part of CSI, e.g., fast-varying small-scale CSI~\cite{Chinacom_2025,10678835}.
Besides, when strict QoS guarantee is required,
extra penalty could be added to $\mathcal{R}_n ( \mathbb{T}_n, \mathbb{T}'_u \,|\, \mathbb{H} )$
to avoid QoS failure~\cite{Chinacom_2025}.
Further, distance definitions different from that in (\ref{eq2}) could be adopted
along with the adjustment of $\mathcal{R}_n ( \mathbb{T}_n, \mathbb{T}'_u \,|\, \mathbb{H} )$ in (\ref{eq1})~\cite{10678835,ref_TWC_2025_submitted}.

Based on (\ref{eq1}) and (\ref{eq2}), the clustering algorithm can be designed for the LAA-satellite links, e.g., using K-means method~\cite{10678835,ref_TWC_2025_submitted}.
Specially, $K$ LAA-satellite link clusters should be formed, one for each carrier.
Depending on whether full or partial frequency reuse is adopted among TBSs,
different link clustering algorithms are needed.
For partial frequency reuse among TBSs, the LAA-satellite links can be hierarchically clustered~\cite{10678835,ref_TWC_2025_submitted}. 
Specifically, coarse link clustering is firstly carried out to 
map LAA-satellite links to the orthogonal carrier subsets formed for partial spectrum sharing among TBSs.
Then fine link clustering is implemented within each coarse cluster to complete scheduling of the links in each carrier.
For instance, when partial frequency reuse with a factor of $4$ is supposed, the LAA-satellite links should be firstly clustered into $4$ coarse clusters.
Coarse and fine LAA-satellite link clusters are respectively indicated by bold and fine dashed circles 
in Fig.~\ref{fig_network_example}.

After the LAA-satellite links are scheduled in $K$ carriers in terms of link clusters,
TBS-TU links can be opportunistically scheduled accordingly.
Instead of relying on link clustering like LAA-satellite links,
scheduling of TBS-TU links can be accomplished based on the solving of assignment subproblems~\cite{10678835,ref_TWC_2025_submitted}.
With appropriate formulation, each subproblem could have polynomial complexity~\cite{10678835}.

\subsection{Link-Feature-Aided Spectrum Sharing with Adaptive Cooperation Time Scale}
\begin{figure} [t]
	\centering
	\includegraphics[width=8.7cm]{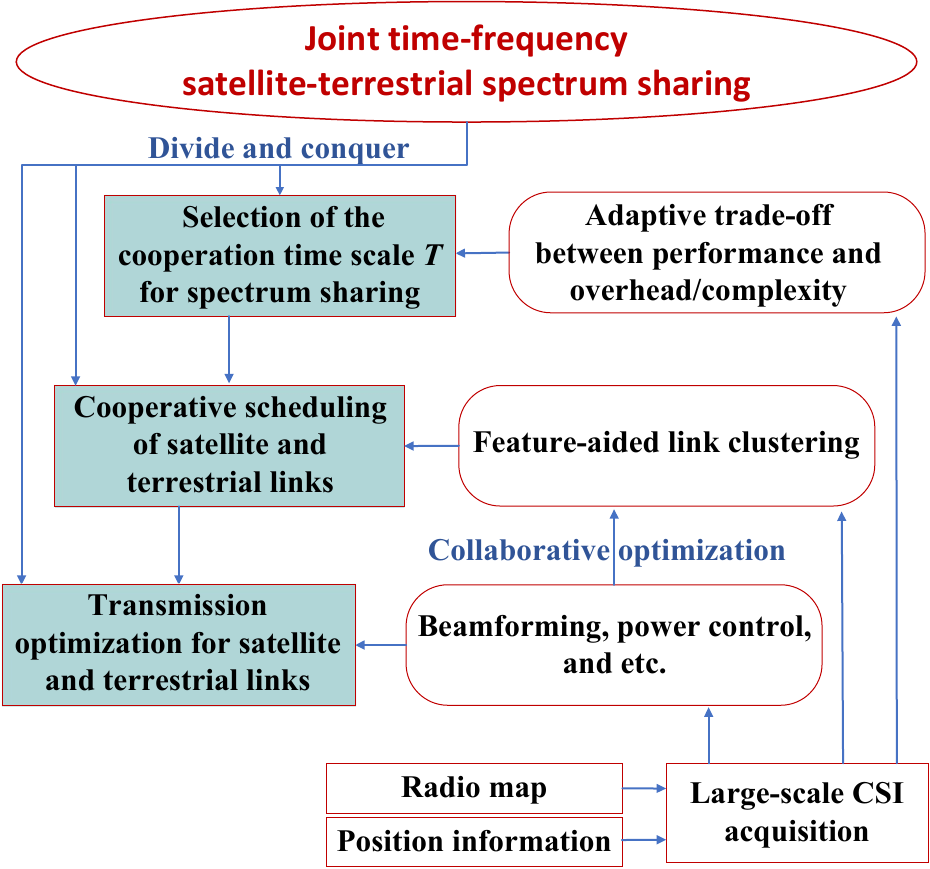}
	\caption{Joint time-frequency satellite-terrestrial spectrum sharing framework.}
	\label{fig_framework}
\end{figure}

Based on feature-aided link clustering, a low-complexity framework can be established for joint time-frequency satellite-terrestrial spectrum sharing via divide-and-conquer. As illustrated in Fig.~\ref{fig_framework}, the framework consists of three main components. First, the value of $T$ is determined through adaptive trade-off between performance and overhead/complexity. Second, cooperative scheduling of satellite and terrestrial links is implemented using feature-aided link clustering. The final step is cooperative transmission optimization, such as beamforming and power control, for satellite and terrestrial links.

Note that by divide-and-conquer,
link scheduling and transmission optimization are decomposed and implemented successively in the framework.
Compared to solving the originally-formulated NP-hard problem directly, performance loss may be incurred.
To minimize the loss,
collaborative optimization should be realized via the decomposed parts.
Specifically, the potential of transmission optimization for the links needs to be taken into account 
when cooperative link scheduling is implemented~\cite{ref_TWC_2025_submitted}. 
As illustrated in the previous subsection,
they can be embedded in the design of feature vectors for link clustering as shown in~(\ref{eq1}).  
For example, suppose power control is utilized together with cooperative link scheduling for spectrum sharing in
the hybrid network.
Then the power control for each pair of LAA-satellite and TBS-TU links can be embedded in 
the design of $\mathcal{R}_n ( \mathbb{T}_n, \mathbb{T}'_u \,|\, \mathbb{H} )$ in~(\ref{eq1})~\cite{ref_TWC_2025_submitted}. 
With only large-scale CSI available for the links,
concise closed-form expressions are often difficult to find for network performance and/or QoS~\cite{10678835,ref_TWC_2025_submitted}.
For example, when average transmission rate is involved 
in the $\mathcal{R}_n ( \mathbb{T}_n, \mathbb{T}'_u \,|\, \mathbb{H} )$ in (\ref{eq1}),
it is challenging to obtain its closed form except in some special cases~\cite{r_JSAC_2021}.
Accordingly,
the Monte Carlo method can be utilized for the evaluation of network performance and/or QoS
in spectrum sharing optimization~\cite{ref_TWC_2025_submitted}.

The large-scale CSI for the links, e.g., $\mathbb{H}$ in~(\ref{eq1}), is critical for spectrum sharing optimization.
It should accurately depict fading statistical characteristics of the satellite, terrestrial and interference links
in each time interval $T$.
Due to diversity of the propagation environment, 
different channel models could be adopted for the links.
Accordingly, the CSI could be represented by distinctive parameters.
Besides, it should match the duration of $T$ as well as movement characteristics of the LAAs and TUs.
Thus, both representation and acquisition of the CSI should be tailored for each specific network scenario,
and even for every link.
Further, as spectrum sharing optimization needs to be completed at the beginning of each time interval,
channel fading prediction is necessary for CSI acquisition.

As large-scale CSI is mainly determined by the positions of transceivers
in a hybrid network with specific propagation environment,
the \emph{radio map} can be utilized for CSI acquisition along with various prediction methods~\cite{r_TVT_2016}.
Different methodologies can be adopted for the construction of radio maps,
based on specialized measurement data, accumulated feedback data of CSI estimation, 
and/or propagation environment information.
To maintain appropriate accuracy, periodic or on-demand update should be carried out to follow the changes of channel fading statistics,
e.g., brought by variation of the propagation environment.
With the aid of the radio map, large-scale CSI of the links can be obtained through table lookup and/or fast calculation
based on positions of the satellites, TBSs, LAAs, and TUs.
Leveraging the inherent position awareness of LAAs, e.g., brought by the pre-planned flight routes, 
it is feasible to introduce the position side information 
into low-altitude-oriented coordinated satellite-terrestrial networks. 
It is also promising to take flight route planning for LAAs into the proposed spectrum sharing framework
for process-oriented optimization across multiple successive time intervals of $T$~\cite{r_JSAC_2021}.
As for the determination of TU positions, enhanced user data collection and reporting can be utilized~\cite{5G_book}.

To effectively grab spectrum sharing opportunities, the accuracy of large-scale CSI needs to be ensured.  
As estimation errors are usually unavoidable, a reasonable criterion should be set for the accuracy 
to balance the performance gain and overhead and complexity.
Robustness of the proposed spectrum sharing framework against estimation errors and outdated radio map data
is a key influencing factor.
While the transmission optimization part of framework can adopt robust algorithms proposed in previous research,
the robustness of link scheduling via feature-aided clustering needs to be further explored. 
Nevertheless, an optimistic result is anticipated
as the main goal of link scheduling is to avoid arranging satellite and terrestrial links
with strong inter-link interference together.
Furthermore, 
robustness of the feature-aided link clustering against CSI errors tends to be vary along with
specific network settings, e.g., the density of TUs and the number of LAAs.
Systematic evaluations can be carried out based on practical network scenarios~\cite{10678835, ref_TWC_2025_submitted}.

\subsection{Complexity and Scalability}
The complexity and scalability of 
feature-aided link clustering is quite critical for the execution efficiency of schemes developed under the framework
in coordinated satellite-5G networks.
Take the network in Fig.~\ref{fig_network_example} as an example.
Basically, the complexity of feature-aided link clustering mainly consists of three polynomial terms, 
i.e., 1) calculation of the feature vectors $\mathcal{F}_u$ for LAA-satellite links
defined in~(\ref{eq1}), 2) clustering of the LAA-satellite links, e.g., via K-means, 
and 3) solving of assignment subproblems for the scheduling of TBS-TU links~\cite{10678835, ref_TWC_2025_submitted}.
The first term is linearly proportional to the number of LAAs and that of TUs.
The second term is linearly proportional to the numbers of LAAs, TUs, available carriers,
and iterations for the clustering algorithm to converge.
The number of available carriers determines that of link clusters to be formed.
Simulation results in our research indicate that a clustering algorithm developed based on K-means
only needs tens of iterations to converge when there are hundreds of TUs and LAAs~\cite{10678835, ref_TWC_2025_submitted}.
The third term grows linearly with the number of TBSs and cubically with that of TUs served by each TBS~\cite{ref_TWC_2025_submitted}.

Note that the complexity analyzed above just acts as a basic benchmark.
In practical applications, there is much potential for it to be reduced.
For example, only inter-link interference above a certain threshold needs to be considered in practice,
and thereby the number of effective elements in $\mathcal{F}_u$ given by~(\ref{eq1}) could be largely decreased.
Besides, TUs with close geographical positions usually lead to similar large-scale channel fading conditions
for both TBS-TU serving links and LAA-TU interference links, and thus could be aggregated into one \emph{virtual TU}
for interference coordination.
Accordingly, the original TUs could be replaced by a much smaller number of virtual TUs in spectrum sharing optimization.
By combing all these measures,
it is rather promising to develop scalable schemes under the proposed framework
even in highly-dense urban scenarios with a massive number of TUs and LAAs.

\section{Application and performance potentials}
\subsection{Potential for Practical Applications}
With an adaptive cooperation time scale $T$ as well as low overhead and complexity,
the spectrum sharing framework is promising for deployment in practical coordinated networks.
By setting appropriate targets or constraints in optimization for the guarantee of network performance and QoS,
different spectrum sharing modes can be realized under the framework. 
Either satellite or terrestrial links are allowed to be the primary side in spectrum sharing,
according to the priority of service they provide~\cite{10678835,ref_TWC_2025_submitted}.
While strict QoS guarantee is usually needed by primary links,
secondary links could provide best-effort service, e.g., for complementary usage~\cite{Chinacom_2025}.

According to practical constraints, 
$T$ can be much larger than one time slot, e.g., for a duration of several or tens of seconds~\cite{10678835,ref_TWC_2025_submitted}. 
Along with the evolution of wireless communications,
either frequency division duplex (FDD) or time division duplex (TDD) can be adopted in the terrestrial component
of the hybrid network~\cite{5G_book}.
When TDD is adopted,
both terrestrial uplinks and downlinks may need to be scheduled in the same carrier in each time interval $T$~\cite{5G_book}.
In that case, terrestrial uplinks and downlinks should be jointly scheduled and optimized 
together with satellite links for spectrum sharing.
Furthermore, the framework can be adapted to evolving transmission technologies.
For example, when massive multiple input multiple output (MIMO) is deployed at the TBSs,
the clustering of terrestrial links could be collaboratively implemented with
the grouping of LAAs/TUs for beam scheduling.

\subsection{Performance Potential by A Case Study}
To explore performance potential of the spectrum sharing framework, a case study is carried out.
Specifically, it is supposed that there is one satellite, orbiting at an altitude of $H_s=500$km, and $M=28$ TBSs.
$U=96$ LAAs flying at the height of $H=200$m are served by the satellite in the uplink via LAA-satellite links.
The coverage area of each TBS is with a radius of $1$km, 
in which $V=24$ uniformly-distributed TUs are served in the downlink via TBS-TU links.
The LAAs are randomly scattered in the airspace above the coverage area of the TBSs following the uniform distribution.
$K=12$ carriers centering at $2$GHz are available, and each of them is with a bandwidth of $1$MHz.
The $U$ LAA-satellite links and $N=MV$ TBS-TU links share the carriers opportunistically
based on a cooperation time scale $T=10$s.
The sub-satellite point is supposed to be located at $(100^\circ E, 40^\circ N)$,
and the center of the coverage area of the TBSs is at $(116^\circ E, 40^\circ N)$.
The receive antenna gain of the satellite is assumed to be $25$dBi, and that of the TUs is $0$dBi.
The transmit antenna gain of the TBSs is $15$dBi, 
and that of the LAAs is determined following the ITU-R recommendation S.465 with a diameter of $0.5$m.
A composite channel model consisting of large-scale and small-scale fading is adopted for the links,
with adaptively-selected parameters~\cite{10678835, ref_TWC_2025_submitted}.

\begin{figure} [t]
	\centering
	\includegraphics[width=8.7cm]{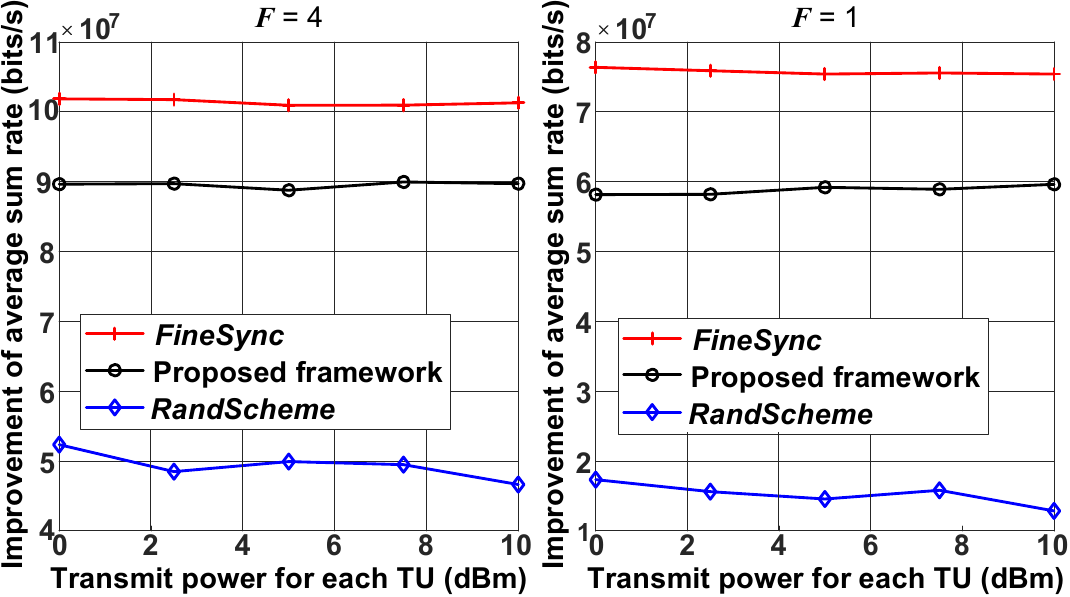}
	\caption{Improvement of the average sum rate of the links.}
	\label{fig_performance}
\end{figure}
 
For generality, both full and partial frequency reuse, with the latter being under a factor of $F=4$, 
are considered for the TBSs~\cite{ref_TWC_2025_submitted}.
To match with the time scale $T$,
the large-scale CSI is set as the path loss of the links, and the shadowing and the small-scale fading 
are supposed to be unknown.
For spectrum sharing,
cooperative scheduling of LAA-satellite and TBS-TU links and power control for LAA-satellite links are implemented.
Without loss of generality,
both LAA-satellite and TBS-TU links are round-robin scheduled via TDMA in the $K$ carriers opportunistically.
The average sum rate of the links is maximized for network performance optimization,
and interference from LAAs to TUs is controlled below a threshold $\gamma_{th}=\sigma^2$[dBm]$-12.2$dB
for QoS guarantee, where $\sigma^2=-114$dBm is the noise power~\cite{ref_TWC_2025_submitted}.
Under the proposed framework,
cooperative link scheduling is accomplished by firstly clustering LAA-satellite links based on interference similarity
and then solving a series of assignment subproblems for scheduling of TBS-TU links~\cite{ref_TWC_2025_submitted}.
The feature vectors $\mathcal{F}_u$ for LAA-satellite links are designed following~(\ref{eq1}).
To take the impact of power control into account,
each $\mathcal{R}_n ( \mathbb{T}_n, \mathbb{T}'_u \,|\, \mathbb{H} )$ in~(\ref{eq1}) is expressed as
a two-element vector. 
The two elements are obtained based on the average transmission rate of the corresponding LAA-satellite link 
under the maximum and minimum allowable transmit power, respectively~\cite{ref_TWC_2025_submitted}.

The improvement of the average sum rate of the links in the $K$ carriers over the case without spectrum sharing,
i.e., only TUs are served via TBS-TU links, is presented in Fig.~\ref{fig_performance}.
The results are obtained by averaging $10$ random topologies of the hybrid network.
The transmit power of the TBSs for each TU is set as $0$dBm to $10$dBm, and that of the LAAs is supposed to be within $[-3,2]$dBW.
For comparison, the performances of two benchmark schemes, i.e., \emph{FineSync} and \emph{RandScheme},
are also presented~\cite{10678835, ref_TWC_2025_submitted}.
Similar to the scheme following the proposed framework, 
\emph{FineSync} is also developed based on cooperative scheduling of LAA-satellite and TBS-TU links and power control for LAA-satellite links~\cite{ref_TWC_2025_submitted}.
The superior point of \emph{FineSync} is that time-slot-level fine satellite-terrestrial synchronization is assumed.
Thus, inter-link interference can be more accurately estimated in the optimization based on specific pairs of LAA-satellite and TBS-TU links
sharing the same carrier simultaneously, instead of just relying on interference profiles derived based on link clusters~\cite{10678835, ref_TWC_2025_submitted}.
\emph{RandScheme} is one with independent satellite-terrestrial link scheduling,
i.e., the LAA-satellite and TBS-TU links are scheduled independently in the carriers.
Besides, no power control is considered for the LAA-satellite links in \emph{RandScheme},
and to avoid severe interference, the transmit power of the LAAs is set as the minimum value, i.e., $-3$dBW.
Fig.~\ref{fig_performance} shows that 
a significant performance gain in spectrum efficiency improvement can be achieved
by the proposed framework~\cite{10678835,ref_TWC_2025_submitted}.
Furthermore,
the gain could be comparative to that with fine satellite-terrestrial synchronization over ms-level time slots.

\section{Related Open Research Directions}
There are lots of open research directions concerning joint time-frequency spectrum sharing
in coordinated satellite-5G networks for low-altitude applications.
Some are as follows.
\begin{itemize}
	\item The planning of flight routes for LAAs can significantly influence spectrum sharing 
		in satellite-5G networks. How to efficiently integrate it with the proposed framework 
		is an interesting issue deserving to be investigated.
	\item The construction, maintenance, and utilization of radio maps for efficient acquisition of large-scale CSI 
	    are preliminarily illustrated in Section IV-C. To support practical applications, more comprehensive study is needed.
	\item Whether radio maps or other prediction methods are utilized for CSI acquisition,
	    deviations from the true values are unavoidable. To set a reasonable criterion for tolerable errors,
	    robustness of feature-aided link clustering should be clarified by further research.
	\item Complicated tasks of LAAs usually last a relatively-long time. Accordingly, transmission for an LAA
		needs to match with the whole process of its task. 
		While the proposed framework is for a single time interval of $T$, task-oriented or process-oriented spectrum sharing strategies 
		across multiple intervals should be further explored.
\end{itemize}

\section{Conclusions}
This paper has addressed joint time-frequency spectrum sharing for hybrid satellite-5G networks serving LAAs. 
To avoid the challenges of fine time synchronization, only coarse synchronization is assumed between satellite and terrestrial links. Specially, cooperative link scheduling is realized via feature-sketching-aided link clustering based on interference similarity, with an adaptive time scale for flexible trade-offs between performance and overhead. 
By utilizing only large-scale CSI and turning to the radio map for CSI acquisition, 
the proposed framework is promising for achieving significant performance gains with low overhead and complexity.
It provides an economically-efficient communication coverage solution for the low-altitude economy.

\begin{IEEEbiographynophoto}{Yanmin Wang} (wangyanmin@muc.edu.cn)
	received the B.S. degree from Shandong University, China, in 2008, and the Ph.D. degree from the Department of Electronic Engineering, Tsinghua University, Beijing, China, in 2013. She is currently an Associate Professor at the School of Information Engineering, Minzu University of China. Her research interests include satellite communications and coordinated satellite-AAV-terrestrial networks.
\end{IEEEbiographynophoto}

\begin{IEEEbiographynophoto}{Wei Feng} (Senior Member, IEEE) (fengwei@tsinghua.edu.cn)
	received the B.S. and Ph.D. degrees from the Department of Electronic Engineering, Tsinghua University, Beijing, China, in 2005 and 2010, respectively. He is currently a Professor with 
	the Department of Electronic Engineering, Tsinghua University. His research interests include 6G
	mobile communications, space-air-ground integrated networks, and Internet of Embodied Al Agents.
\end{IEEEbiographynophoto}

\begin{IEEEbiographynophoto}{Yunfei Chen} (Fellow, IEEE) (yunfei.chen@durham.ac.uk)
	received the B.E. and M.E. degrees in electronics engineering from Shanghai Jiaotong University, Shanghai, China, in 1998 and 2001, respectively, and the Ph.D. degree from the University of Alberta in 2006. He is currently a Professor with the Department of Engineering, University of Durham, U.K. His research interests include wireless 
	communications, performance analysis, and joint radar communications designs.
\end{IEEEbiographynophoto}

\begin{IEEEbiographynophoto}{Baoquan Ren} (renbq88@126.com)
	received the Ph.D. degree from the
	Institute of China Electronic System Engineering
	Corporation, Beijing, China, in 2010.
	He is currently a Senior Engineer with China Electronic System Engineering Company, Beijing. His research interests
	include Internet of Things and wireless communication technologies.
\end{IEEEbiographynophoto}

\begin{IEEEbiographynophoto}{Qingqing Wu} (Senior Member, IEEE) (qingqingwu@sjtu.edu.cn)
	 is a Professor with Shanghai Jiao Tong University. His current research interests include intelligent reflecting surface (IRS), uncrewed aerial vehicle (UAV) communications, and MIMO transceiver design. He has co-authored over 100 IEEE journal papers, which have received over 50000 Google Scholar citations. He has been listed as a Clarivate ESI Highly Cited Researcher since 2021. 
\end{IEEEbiographynophoto}

\begin{IEEEbiographynophoto}{Cheng-Xiang Wang} (Fellow, IEEE) (chxwang@seu.edu.cn)
received the Ph.D. degree in wireless communications from Aalborg University, Denmark, in 2004. He is currently a Professor with the National Mobile Communications Research Laboratory, Southeast University, China and a Vice President of Southeast University. He has authored 5 books and over 390 papers in refereed journals, including 37 highly cited papers. His research interests include wireless channel measurements and modeling, 6G/B6G intelligent wireless networks, and electromagnetic information theory. He is a member of the Academia Europaea.
\end{IEEEbiographynophoto}

\vfill

\end{document}